\documentclass[aps,prb,twocolumn,amsmath,amssymb,superscriptaddress]{revtex4}
\usepackage{graphicx}
\usepackage{bm}
\usepackage{mathptmx}
\usepackage[english]{babel}
\usepackage{indentfirst}
\usepackage{color}
\usepackage{listings}
\usepackage{float}

\usepackage{hyperref}

\begin{document}

\title{Topological edge state engineering with high-frequency electromagnetic radiation}

\author{Mehedi Hasan}
\affiliation{ITMO University, Saint Petersburg 197101, Russia}
\affiliation{Division of Physics and Applied Physics, Nanyang Technological University 637371, Singapore}

\author{Dmitry Yudin}
\affiliation{ITMO University, Saint Petersburg 197101, Russia}

\author{Ivan Iorsh}
\affiliation{ITMO University, Saint Petersburg 197101, Russia}
\affiliation{Division of Physics and Applied Physics, Nanyang Technological University 637371, Singapore}

\author{Olle Eriksson}
\affiliation{Department of Physics and Astronomy, Uppsala University, Box 516, SE-751 20 Uppsala, Sweden}
\affiliation{School of Science and Technology, \"Orebro University, SE-701 82 \"Orebro, Sweden}

\author{Ivan Shelykh}
\affiliation{ITMO University, Saint Petersburg 197101, Russia}
\affiliation{Science Institute, University of Iceland IS-107, Reykjavik, Iceland}

\begin{abstract}
We outline here how strong light-matter interaction can be used to induce quantum phase transition between normal and topological phases in two-dimensional topological insulators. We consider the case of a HgTe quantum well, in which band inversion occurs above a critical value of the well thickness, and demonstrate that coupling between electron states and the $E$ field from an off-resonant linearly polarized laser provides a powerful tool to control topological transitions, even for a thickness of the quantum well that is below the critical value. We also show that topological phase properties of the edge states, including their group velocity, can be tuned in a controllable way by changing the intensity of the laser field. These findings open up the possibility for new experimental means with which to investigate topological insulators and shed new light on topological-insulator-based technologies that are under active discussion.
\end{abstract}

\maketitle

\section{Introduction}
Over the last few decades, the study of topological properties of matter and emergent phenomena has evolved into a mature field: topological ideas are widely used in condensed-matter physics \cite{Hasan2010,Qi2011}, photonics \cite{Lu2014}, ultracold quantum gases \cite{Goldman2016}, and polaritonics \cite{Karzig2015}. It turns out that the nontrivial topology of the bulk band structure results in the formation of gapless edge modes, which makes the surface or interface of the material conducting while leaving the interior insulating. Being robust against smooth variations of the material parameters that are relevant for the Hamiltonian, these edge modes are topologically protected and can be destroyed only by closing the bulk gap. Well-known examples of these topologically protected states are the edge states emerging in the quantum Hall phase \cite{Thouless1982,Hatsugai1990} and topological insulators \cite{Kane2005a,Kane2005b,Bernevig2006}. 

Soon after the theoretical prediction \cite{Zhang2006}, a $\mathbb{Z}_{2}$ topological insulator was realized in a HgTe/CdTe quantum well \cite{Konig2007}, where the existence of edge states has been clearly demonstrated experimentally by measuring quantized resistance. The parameter which conventionally governs the transition to the topological state is the thickness of the HgTe layer: below a certain critical value (approximately 6.1 nm) the system is in the normal state, while above it, band inversion occurs and the transition to a topological state takes place \cite{Zhang2006}. Clearly, once the structure is grown, the thickness of the layer cannot be changed, and thus, transitions between topological and normal states in the same sample are irreversible. It will be demonstrated later in this paper that this transition can be realized in a reversible manner if one exploits an additional element of control, namely, coupling of the system with an external electromagnetic wave. 

In the most general case, the problem of a quantum structure interacting with an external time-dependent field has no analytic solution and cannot be characterized by well-defined quantum numbers, such as particle energy. Nevertheless, the case of a time-periodic external field can be addressed by means of the so-called Floquet expansion \cite{Grifoni1998,Kohler2005}, which allows the description of the system in terms of quasistationary Floquet states characterized by Floquet quasienergies. The topological classification of periodically driven systems has been proposed in Ref. \cite{Kitagawa2010}, where it was shown that the dynamical character of a driving field gives rise to a variety of topological phases akin to time-independent ones. For the topological state created with an external time-periodic perturbation, the term Floquet topological insulator has been coined, as it exhibits edge states in the gap of its quasienergy spectrum. In these systems, the periodic field drives the system into a nonequilibrium topological state, and topological properties can be induced even in conventional band insulators \cite{Oka2009,Inoue2010,Karch2010,Lindner2011,Gu2011,Kitagawa2011,Calvo2011,Dora2012,Morell2012,Rechtsman2013,Wang2013,Katan2013,Iadecola2013,Leon2013,Liu2013,Rudner2013,Kundu2013,Fregoso2013,Cayssol2013,Jotzu2014,Piskunow2014,Grushin2014,Usaj2014,Dehghani2014,Kundu2014,Wang2014,Sentef2015,Titum2015,Dehghani2015,Dahlhaus2015,Klinovaja2016,Farrell2016}. Meanwhile, the model of the Floquet topological insulator has been considered since the seminal work \cite{Lindner2011}, and it exhibits a single pair of helical edge modes due to the on-resonant light-induced band inversion. When irradiating with an on-resonant external light, the bands in HgTe/CdTe are reshuffled. Thus, the effective Hamiltonian undergoes band inversion, subsequently allowing for the topological insulator to be realized in a broad range of photon frequencies. For on-resonant light, in the bulk spectrum there is an anticrossing separating the reshuffled valence band from the conduction band.

In contrast to what has been considered so far in the context of the Floquet topological insulator, in this paper we demonstrate how the coupling to off-resonant, linearly polarized electromagnetic radiation modifies the electronic properties of a $\mathbb{Z}_{2}$ topological insulator in a nonintuitive, yet useful, way. For light with off-resonant frequency, the real process of photon absorption or emission cannot occur because of the constraints imposed by the energy conservation. However, the off-resonant light can affect the electron system via virtual processes \cite{Ohtsu2013}. We show here that the light-matter coupling in a $\mathbb{Z}_{2}$ topological insulator leads to a significant renormalization of the parameters of the Hamiltonian that describe the electronic structure while preserving time-reversal symmetry ($T$ symmetry). In the most interesting scenario, this renormalization is demonstrated to drive a transition between normal and topological phases, which strongly modifies the properties of the edge states, in a controllable and reversible manner. The rest of the paper is organized as follows: in Sec.~\ref{sec:expansion} we apply high-frequency expansion in the form of the Brillouin-Wigner perturbation theory to the Hamiltonian of the HgTe quantum well. We further discuss the results of our numerical simulations, revealing the formation of edge states owing to light-matter interaction and the possibility to tune their properties by adjusting parameters of the driving, in Sec.~\ref{sec:results}. Finally, we summarize our findings in Sec.~\ref{sec:conclusions}

\section{High-frequency expansion}\label{sec:expansion}
\begin{figure}[htpb]
\centering
\includegraphics[scale=0.15]{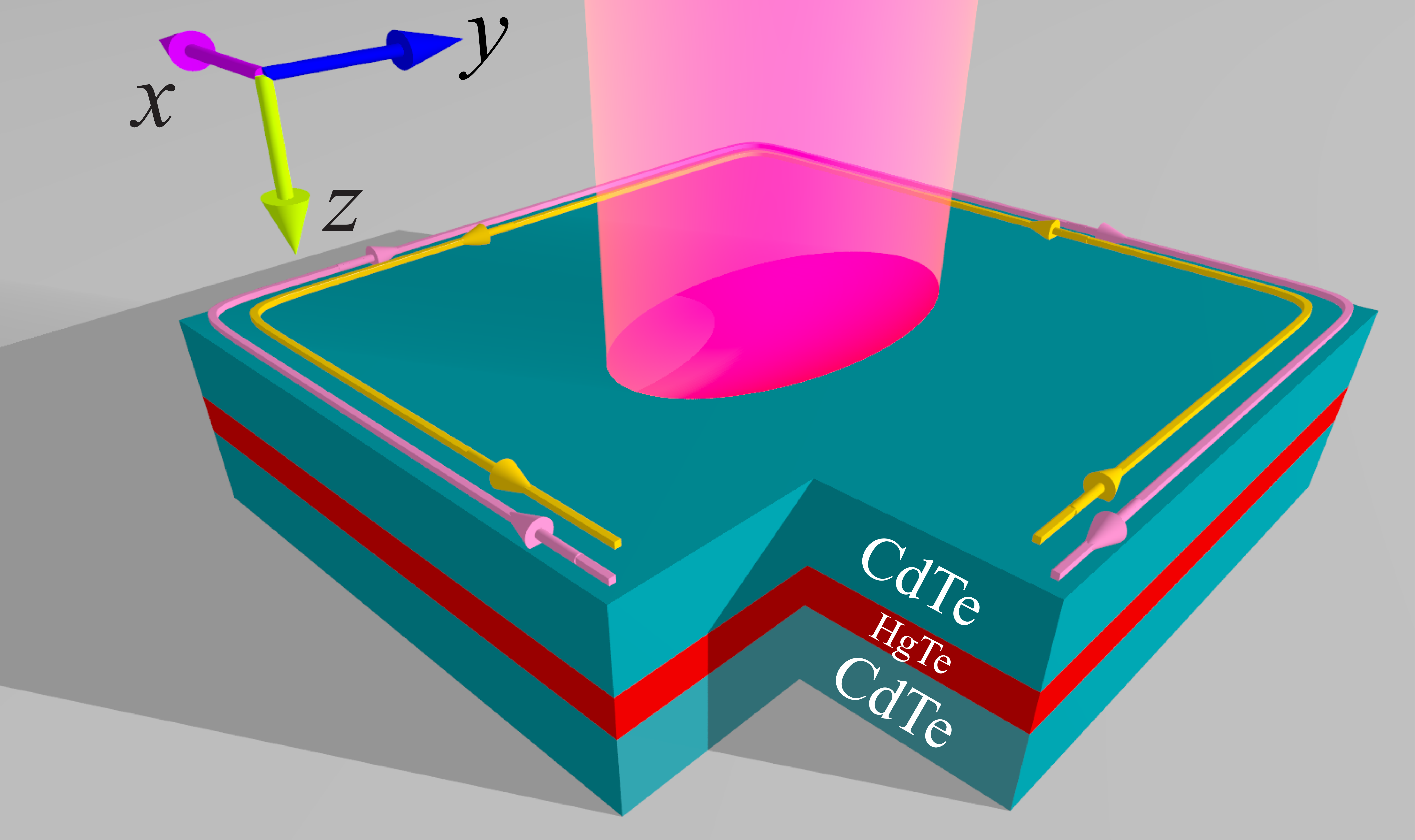}
\caption{Geometry of the system under consideration. A HgTe quantum well is sandwiched between CdTe layers and is pumped with high-intensity off-resonant electromagnetic field (the purple beam) normal to its interface. The coupling between electrons and electromagnetic field modifies the parameters of the Hamiltonian in Eq.~(\ref{inham1}), and this leads to the transition between normal and topological phases, and the control of edge-state properties.}
\label{fig:fig1}
\end{figure}

In the following, we develop a theoretical description of a Floquet topological insulator in the zinc-blende structure, where we consider the Hamiltonian, appropriate for a HgTe quantum well coupled to a linearly polarized electromagnetic wave that propagates perpendicular to the quantum well interface, as shown in Fig.~\ref{fig:fig1} schematically. We assume a negligible time dependence of the amplitude of the field, which allows us to work in the paradigm of the adiabatic approximation.

The effective electronic Hamiltonian of a HgTe quantum well sandwiched between CdTe layers can be derived from the group-theoretic method of invariants or obtained by fitting parameters of the Hamiltonian to results of {\it ab initio} calculations. In the vicinity of the $\Gamma$ point it can be written as \cite{Zhang2006,Gerchikov1989,Tarasenko2015,Durnev2016}
\begin{equation}\label{inham1}
H=-dk^2+\left(\begin{array}{cc}
h_\mathbf{k} & i\gamma\sigma_x \\
-i\gamma\sigma_x & h_\mathbf{k}^T
\end{array}\right),
\end{equation}
where $h_\mathbf{k}=(\delta_0-bk^2)\sigma_z-a\left(k_y\sigma_x+k_x\sigma_y\right)$, $\bm{\sigma}$ is a vector of Pauli matrices, and a two-dimensional wave vector $\mathbf{k}=\left(k_x,k_y\right)$ specifies electron states. In general, $\delta_0$ corresponds to the gap, while the parameters $a$, $b$, and $d$ describe the band dispersion, and as mentioned above, their numerical values may be obtained from first-principles calculations or from experimental data \cite{Zhang2006,Konig2008}. The parameter $\gamma$ is purely determined by the interface asymmetry of the quantum well, and in the special case when $\gamma=0$, the Hamiltonian in Eq.~(\ref{inham1}) coincides with that of Bernevig, Hughes, and Zhang \cite{Zhang2006}. However, $\gamma$ has recently been shown to have a dramatic impact on the properties of quantum wells, as it drives a rather strong level repulsion which drastically modifies the electron states and dispersion \cite{Tarasenko2015}. Hence, a nonvanishing value should be kept to make theoretical predictions that are relevant for experimental observations.

The geometry we consider throughout the calculations is schematically depicted in Fig.~\ref{fig:fig1}. A linearly polarized electromagnetic wave irradiates the quantum well and is assumed to be uniform in the plane of the well. The time dependence is induced via minimal coupling of electron states to the electromagnetic vector potential: $\hbar\mathbf{k}\rightarrow\hbar\mathbf{k}-e\mathbf{A}(t)$, where $\mathbf{A}(t)=\mathbf{E}_0\cos\omega t/\omega$ and the field is chosen to be polarized along the $y$ axis, i.e., $\mathbf{E}_0=E_0\mathbf{e}_y$ (the choice of coordinate system is shown in Fig.~\ref{fig:fig1}). To observe tangible effects in experiments, the frequency $\omega$ of the external driving field has to be tuned to avoid optical excitations. In this situation, no real absorption or emission of photons by the electronic subsystem can happen. However, virtual processes may drastically modify the corresponding band structure, as shown below. For the sake of simplicity, we put $d=0$, which corresponds to a case with particle-hole symmetry. This choice simplifies the analytical calculation while still being realistic.
\begin{figure*}[ht!]
\centering
\includegraphics[scale=0.21]{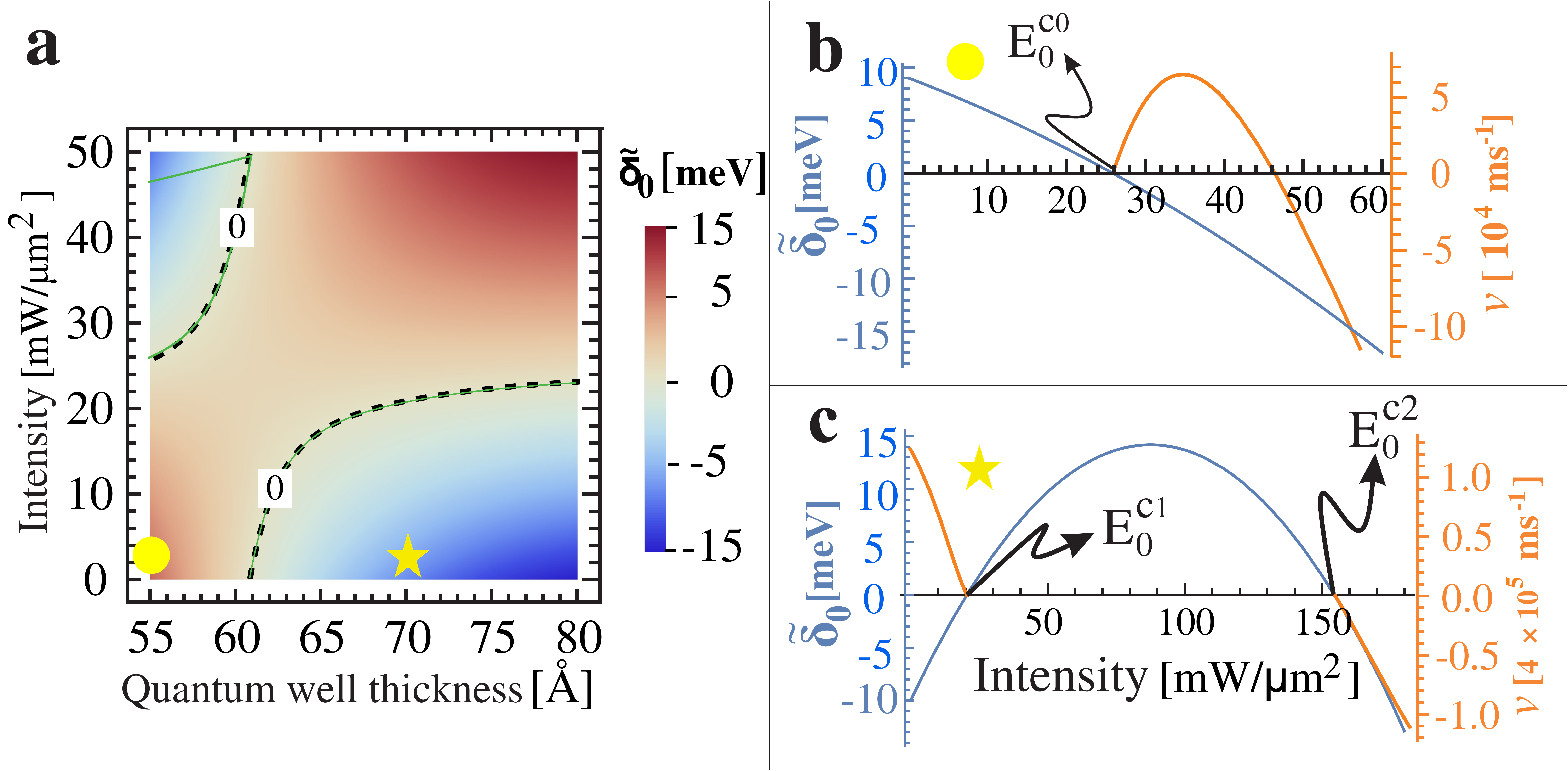}
\caption{Light-induced phase transitions under off-resonant pumping with electromagnetic radiation. (a) The phase diagram of the system, revealing transitions between topologically trivial and nontrivial states depending on quantum well thickness and the intensity of the light. The dashed lines stand for $\tilde{\delta}_0=0$ and separate the topological areas (bottom right and top left corners with $\tilde{\delta}_0<0$) from the band-insulating one (the central region with $\tilde{\delta}_0>0$). The green lines in (a) mark the region with zero group velocity of the edge states. There exist two types of phase transitions as a function of the intensity of the electromagnetic field: (b) band insulator to topological insulator and (c) topological insulator to band insulator to topological insulator. The blue line and left scale show the dependence of the renormalized gap parameter $\tilde{\delta}_0$ on the pump intensity. The orange line and right scale correspond to the group velocity of the edge states in the topological phase. 
In the calculations we assume the quantum well thickness is chosen as $l=5.5$ nm in (b), with $\delta_0=9$ meV, $a=3.87$ eV \AA, $b=-48$ eV \AA$^2$, and $l=7$ nm in (c), with $\delta_0=-10$ meV, $a=3.65$ eV \AA, $b=-68.6$ eV \AA$^2$.}
\label{fig:fig2}
\end{figure*}

In the high-frequency regime of external pumping, Floquet bands become almost uncoupled, thus making the Hamiltonian block diagonal in Fourier space. The latter permits us to fill the bands akin to the time-independent systems without any need to estimate the electronic occupancy of Floquet sidebands. The rigorous derivation of the effective time-independent Hamiltonian of quantum systems driven by a high-frequency pumping basically represents a perturbative expansion with respect to $1/\omega$. A formal mathematical routine may be based on one of three alternatives: (i) Floquet-Magnus expansion \cite{Magnus1954,Casas2001,Blanes2009,Mananga2011}, which attempts to write down a stroboscopic evolution operator for a time-independent effective Hamiltonian, (ii)  the van Vleck perturbative series \cite{Eckardt2015,Bukov2015}, employing a special ansatz for the time-evolution operator, and (iii) Brillouin-Wigner perturbation theory, which was recently proposed for this class of problems \cite{Mikami2016} within which one projects an extended Hilbert to a zero-photon subspace. The advantage of utilizing the Brillouin-Wigner perturbation theory, which we use in the following, is that in contrast to Floquet-Magnus expansion, which explicitly depends on the phase of a driving field, and the van Vleck approach, which generates an unlimited number of terms already in the lowest orders of perturbative expansion, it is computationally efficient and correctly reproduces well-known results of the energy spectrum. The first nonvanishing correction to Eq.~(\ref{inham1}), stemming from the terms proportional to $1/\omega^2$, leads to
\begin{equation}\label{finham}
\tilde{H}=\left(\begin{array}{cc}
\tilde{h}_\mathbf{k} & i\tilde{\gamma}\sigma_x \\
-i\tilde{\gamma}\sigma_x & \tilde{h}^T_\mathbf{k}
\end{array}\right),
\end{equation}
where $\tilde{h}_\mathbf{k}=(\tilde{\delta}_0-B_xk_x^2-B_yk_y^2)\sigma_z-(A_xk_x\sigma_y+A_yk_y\sigma_x)$. Thus, placing the quantum well in a strong electromagnetic field results in a renormalization of $\delta_0$, $\gamma$, $a$, $b$, and $d$ (the corresponding expressions of $\tilde{\delta}_0$, $\tilde{\gamma}$, $A_x$, $A_y$, $B_x$, and $B_y$ are listed in Appendix~\ref{appendix:a}) in such a way that the final Hamiltonian [Eq.~(\ref{finham})] becomes highly anisotropic in $\mathbf{k}$ space. It is noteworthy that the effective Hamiltonian in Eq.~(\ref{finham}) still respects $T$ symmetry, thus guaranteeing the existence of topologically protected edge states \cite{Hasan2010,Qi2011}. This is in sharp contrast to the topological transitions induced by circularly polarized light (see, e.g., ~\cite{Oka2009}), as the latter serves as an effective magnetic field: it breaks $T$ symmetry, thus leading to the gap opening and destroying topological protection (the explicit expression for the effective Hamiltonian for the circularly polarized radiation is derived in Appendix~\ref{appendix:b}).

The topologically nontrivial state corresponds to negative values of $b$ and $\delta_0$ of the original Hamiltonian (\ref{inham1}) \cite{Qi2011}, while the influence of external pumping on the gap is determined by its renormalized value $\tilde{\delta}_0$,
\begin{equation}\label{rengap}
\tilde{\delta}_0=\delta_0-\left(\frac{a^2\delta_0}{2\hbar^2\omega^2}+\frac{b}{2}\right)\left(\frac{eE_0}{\hbar\omega}\right)^2 +\frac{\left(b\delta_0+8a^2\right)b}{32\hbar^2\omega^2}\left(\frac{eE_0}{\hbar\omega}\right)^4.
\end{equation}
It is clear therefrom that by tuning the intensity ($\sim E_0^2$) and frequency of the electromagnetic radiation, one can change the sign of $\tilde{\delta}_0$, something which allows us to drive topological transitions in the system solely by optical means. This is the central result we would like to convey in this paper: dressing by an off-resonant linearly polarized electromagnetic field renormalizes the parameters of the Hamiltonian in such a way that phase transitions between topologically trivial and nontrivial states can be realized, in an easily controllable and reversible manner.

\section{Results and discussion}\label{sec:results}
To illustrate our analytical findings we present in Fig.~\ref{fig:fig2}(a) a phase diagram that describes the transition between topological and band-insulating phases depending on quantum well thickness and light intensity. The increase of the intensity of the electromagnetic radiation exposing the quantum system on the topologically trivial side (marked by a circle) leads to a change of the sign of $\tilde{\delta}_0$ (the dashed black line). The system undergoes the phase transition to the topologically insulating state [top left corner of Fig.~\ref{fig:fig2}(a)]. However, as shown in Fig.~\ref{fig:fig2}(b), the group velocity $v$ (orange curve, right axis) is positive at the onset of the topological phase-transition point $E_0^{c0}$, and it changes sign as the intensity is increased. This suggests that the group velocity of opposite-propagating spins will flip each of their propagation directions (Appendix~\ref{appendix:c}). On the other hand, for a system originally on the topologically nontrivial side (marked by a star) the increase of the intensity of external pumping drives the transition to normal at the point $E_0^{c1}$ and then back to the topological phase at $E_0^{c2}$ [see Fig.~\ref{fig:fig2}(c)]. In the latter case the spin currents propagate in opposite directions for the two different topological phases. Throughout the calculations in Fig.~\ref{fig:fig2}, we work with the frequency $\omega=60$ THz, and other parameters relevant for the calculation are listed in the caption of Fig.~\ref{fig:fig2}.

\begin{figure}[H]
\centering
\includegraphics[scale=0.2]{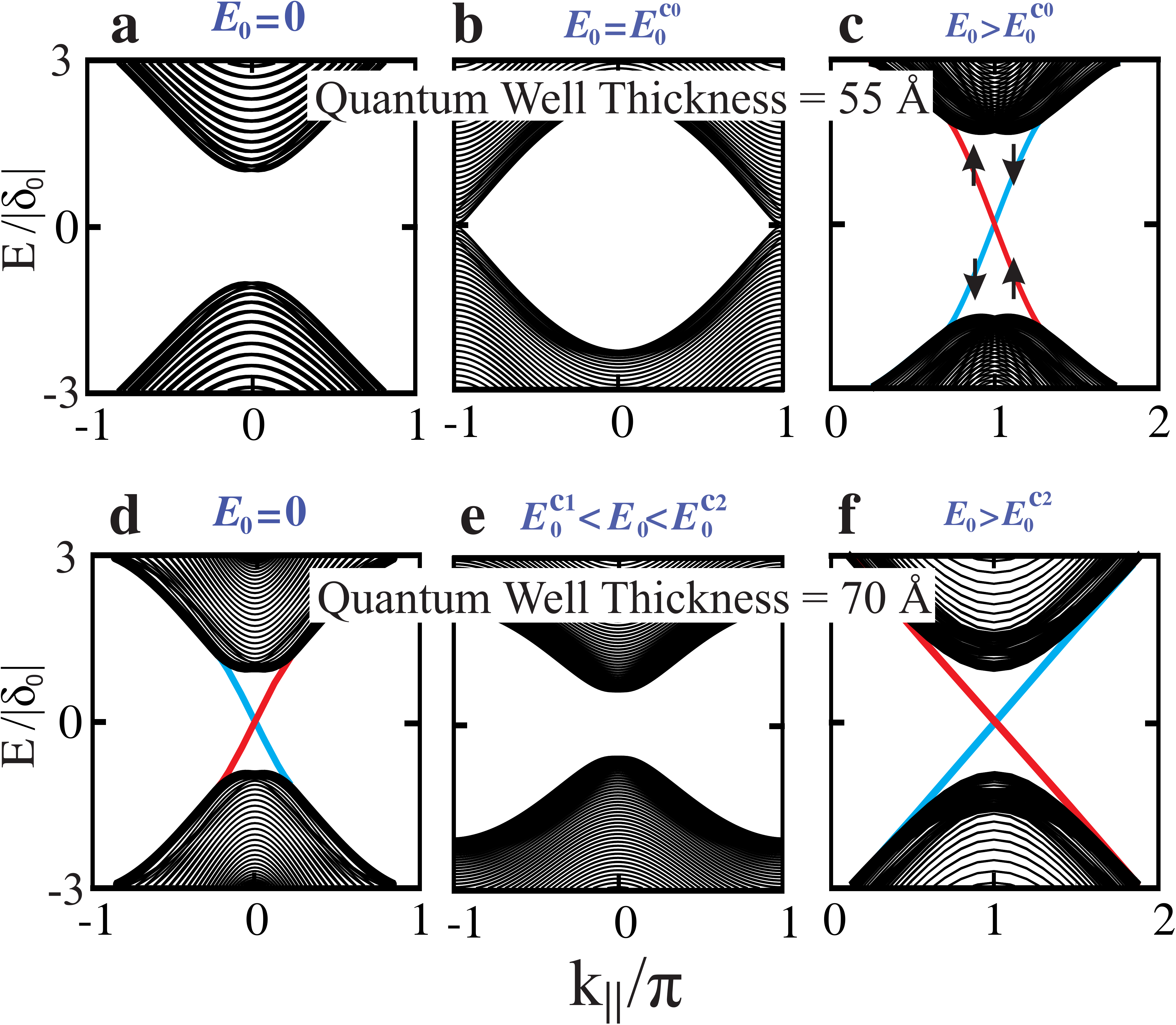}
\caption{Band diagram for the finite strip of the CdTe/HgTe/CdTe quantum well (the energy $E$ is measured in units of $\vert\delta_0\vert$, and $k_\parallel$ is the wave vector along the edge; the order of magnitude of $\delta_0$ is meV, as seen from Fig.~\ref{fig:fig2}). (a)--(c) The evolution of a band insulator to topological insulator: (a) in the absence of radiation the system is in a band-insulating phase.  The increase of the radiation field leads to (b) the quenching of the gap and, eventually, (c) a transition to the topological phase, when edge states, marked by red (spin up) and blue (spin down) lines, show up. (d)--(f) Phase transition from a topological insulator to band insulator to topological insulator: (d) in the absence of the field edge states are present (red and blue lines). (e) As the intensity of the radiation goes up the system undergoes a transition to a normal phase, and the edge states disappear. (f) A further increase of the intensity of the light causes a transition back to the topological phase. Note that directions of spin currents in (d) and (f) are opposite, as regimes of low and high pump correspond to the different signs of the group velocity [see Fig.~\ref{fig:fig2}(c)].}
\label{fig:fig3}
\end{figure}

The effect of external pumping turns out to have a profound impact on the properties of the edge states of the renormalized Hamiltonian (\ref{finham}). Results of our numerical simulations reveal their formation in a finite strip of the material (see Appendix~\ref{appendix:d} for a description of the numerical calculations). For a band insulator [Fig.~\ref{fig:fig3}(a)] the increase in the laser intensity gives rise to the closing of the gap at the critical value of the field, $E_0=E_0^{c0}$ [Fig.~\ref{fig:fig3}(b)], while the subsequent increase of $E_0$ is found to cause band inversion and a transition to the topological phase in which edge states appear [Fig.~\ref{fig:fig3}(c)]. 

Otherwise, for a structure initially in a topological phase, edge states are present [Fig.~\ref{fig:fig3}(d)], and an increase in the field strength to a value $E_0=E_0^{c1}$ drives the system first into a normal phase, where edge states are absent [Fig.~\ref{fig:fig3}(e)], and then back to the topological state, where they reappear [Fig.~\ref{fig:fig3}(f)]. Note that the sign of the group velocity is different in the two topological phases, which physically corresponds to the opposite directions of the propagation of edge spin currents.

\section{Conclusions and outlook}\label{sec:conclusions}

To summarize, in this work we have analyzed how the interaction between electrons and an external electromagnetic field modifies strongly the electronic properties of a $\mathbb{Z}_2$ topological insulator. We considered the classical example of the CdTe/HgTe/CdTe heterostructure and demonstrated that coupling between electrons in this quasi-two-dimensional material and an external electromagnetic field can drive transitions between normal and topological phases in a reversible manner. Depending on the thickness of the quantum well, i.e., the initial state of the system, we find transitions from a normal state to a topological state or transitions from a topological state to a normal state and back to a topological state as the intensity of the electromagnetic radiation increases. We have further shown that the light-matter interaction modifies the properties of the edge states in a quite controllable fashion. In particular, this leads to the creation and control over the magnitude and directions of the edge spin currents. 

To discuss the possible verification of our theoretical findings, we provide a simple quantitative estimate of the parameters of an experimental setup. The off-resonant character of light-matter coupling is dictated by the frequency of the field to be tuned to avoid optical excitations of electrons, i.e., $\omega\sim\mathrm{max}\{a/\Delta,b/\Delta^2\}$ (where $\Delta\sim$10 $\AA$ is the corresponding lattice spacing). The latter can be achieved for $\omega\sim10^2-10^3$ THz; meanwhile, the results of our numerical simulations reveal that qualitative behavior does not change even for lower frequencies (see Appendix~\ref{appendix:e} for further details). We expect that recent experimental progress \cite{Wang2013} in time-resolved angle-resolved photoemission spectroscopy would enable direct verification of our results, while complementary transport measurements (e.g., Hall conductivity) would provide a versatile tool for testing the validity of the static Hamiltonian approximation.

Our findings open up new experimental means to investigate topological insulators. Importantly, all the topological effects discussed here can be realized in a controllable and reversible manner. This has an impact on the development of nanotechnological devices, concerning both prospective spin-optical and electron-optical devices, for instance, in the form of transistors, where the electromagnetic field would act as a gate that turns ``off'' or ``on'' the conducting edge states shown in Fig.~\ref{fig:fig1}. Other technologies that may evolve from our findings relate to the predicted, controllable group velocity of the edge states. As Fig.~\ref{fig:fig2}(b) shows, by tuning the intensity of the electromagnetic radiation, the electrons at the edge can be made to move forwards or backwards or to be conducting but with an essentially vanishing speed. 

\section*{Acknowledgements}

Valuable discussions with Dr. P. Belov are acknowledged. We acknowledge the support from the Russian Academic Excellence Project ``5-100'', from Projects No. 3.1365.2017/4.6, No. 3.8884.2017/8.9, No. 3.4424.2017/HM, and No. 3.2614.2017/4.6 of the Ministry of Education and Science of the Russian Federation, from Megagrant 14.Y26.31.0015, and from the Icelandic research fund (Rannis) under Projects No. 163082-051. D.Y. acknowledges the support from RFBR Project No. 16-32-60040. I.I. acknowledges support from RFBR Project No. 16-32-60123. O.E. acknowledges support from the Swedish Research Council, the KAW foundation (Grants No. 2012.0031 and No. 2013.0020), and eSSENCE. I.S. acknowledges the support from Horizon2020 projects NOTEDEV and CoExAN.

\appendix

\section{Derivation for linearly polarized pumping}\label{appendix:a}

It was mentioned in the main text that in the vicinity of the $\Gamma$ point, specified by a two-dimensional wave vector $\mathbf{k}=\left(k_x,k_y\right)$, the Hamiltonian of a HgTe quantum well reads 
\begin{equation}\label{inham}
H=\left(\begin{array}{cc}
h_\mathbf{k} & i\gamma\sigma_x \\
-i\gamma\sigma_x & h^T_\mathbf{k}
\end{array}\right),
\end{equation}
where we have defined $\sigma=\left(\sigma_0,\bm{\sigma}\right)$ as a set of Pauli matrices equipped with a $2\times2$ unity matrix. The $2\times2-$blocks of the Hamiltonian $h_\mathbf{k}$ and $h_\mathbf{k}^T$ act on the subspace spanned by the states with total momentum $J_z=1/2,3/2$ and $J_z=-1/2,-3/2$, respectively, and they are related to each other by time-reversal symmetry. An explicit expression for $h_\mathbf{k}$ is
\begin{equation}\label{inham2}
h_\mathbf{k}=\left(\begin{array}{cc}
\delta_0-(b+d)k^2 & ia(k_x-ik_y) \\
-ia(k_x+ik_y) & -\delta_0+(b+d)k^2
\end{array}\right).
\end{equation}
The time-dependence $H(t)=\sum H_n\,e^{in\omega t}$ is added to the initial Hamiltonian (\ref{inham}) via the electromagnetic vector potential of the external driving field, $\mathbf{E}(t)=\mathbf{E}_0\sin\omega t$, with frequency $\omega$ (this is assumed to be the dominant energy scale) and with a polarization along the $y$ axis: $\mathbf{E}_0=E_0\mathbf{e}_y$. Next, we define
\begin{equation}\label{fourier0}
H_0=H-\frac{b}{2}\left(\frac{eE_0}{\hbar\omega}\right)^2\left(\begin{array}{cc}
\sigma_z & 0 \\
0 & \sigma_z
\end{array}\right),
\end{equation}
\begin{equation}\label{fourier1}
H_1=H_{-1}=\left(\frac{eE_0}{\hbar\omega}\right)\left(\begin{array}{cc}
bk_y\sigma_z+\frac{a}{2}\sigma_x & 0 \\
0 & bk_y\sigma_z+\frac{a}{2}\sigma_x
\end{array}\right),
\end{equation}
\begin{equation}\label{fourier2}
H_2=H_{-2}=\frac{b}{4}\left(\frac{eE_0}{\hbar\omega}\right)^2\left(\begin{array}{cc}
\sigma_z & 0 \\
0 & \sigma_z
\end{array}\right).
\end{equation}
For the sake of simplicity, and as mentioned in the main text, we put $d=0$, which corresponds to electron-hole symmetry. The effective renormalized Hamiltonian can be derived within the paradigm of high-frequency expansion in the form of Brillouin-Wigner theory. Up to the terms in $1/\omega^2$ we obtain
\begin{equation}
\tilde{H}=H+\frac{1}{(\hbar\omega)^2}\left\lbrace\left[H_1,\left[H_1,H_2\right]\right]+2\sum\limits_{n=1,2}\frac{H_nH_0H_n}{n}\right\rbrace.
\end{equation}
After straight forward algebra we derive
\begin{equation}\label{effham}
\tilde{H}=\left(\begin{array}{cc}
\tilde{h}_\mathbf{k} & i\left(\tilde{\gamma}+\gamma^\prime_yk_y^2\right)\sigma_x \\
-i\left(\tilde{\gamma}+\gamma^\prime_yk_y^2\right)\sigma_x & \tilde{h}^T_\mathbf{k}
\end{array}\right),
\end{equation}
where the $2\times2$ block
\begin{equation}\label{effham2}
\tilde{h}_\mathbf{k}=\left(\begin{array}{cc}
\tilde{\delta}_0-B_xk_x^2-B_yk_y^2 & i(A_xk_x-iA_yk_y) \\
-i(A_xk_x+iA_yk_y) & -\tilde{\delta}_0+B_xk_x^2+B_yk_y^2
\end{array}\right).
\end{equation}
is formally analogous to $h_\mathbf{k}$ with renormalized parameters. In formula (\ref{effham}) we have defined renormalized parameters:
\begin{equation}\label{gamma}
\frac{\tilde{\gamma}}{\gamma}=1+\frac{a^2}{2\hbar^2\omega^2}\left(\frac{eE_0}{\hbar\omega}\right)^2-\frac{b^2}{32\hbar^2\omega^2}\left(\frac{eE_0}{\hbar\omega}\right)^4.
\end{equation}
It is worthwhile to note that such a procedure generates $\gamma_y^\prime=\frac{2\gamma b^2}{\hbar^2\omega^2}\left(\frac{eE_0}{\hbar\omega}\right)^2$, which is present in any realistic quantum well structure owing to interface mixing asymmetry. The rest correspond to
\begin{equation}\label{delta}
\tilde{\delta}_0=\delta_0-\left(\frac{a^2\delta_0}{2\hbar^2\omega^2}+\frac{b}{2}\right)\left(\frac{eE_0}{\hbar\omega}\right)^2+\frac{\left(b\delta_0+8a^2\right)b}{32\hbar^2\omega^2}\left(\frac{eE_0}{\hbar\omega}\right)^4,
\end{equation}
i.e., a renormalized gap, and
\begin{equation}\label{ax}
\frac{A_x}{a}=1+\frac{a^2-4b\delta_0}{2\hbar^2\omega^2}\left(\frac{eE_0}{\hbar\omega}\right)^2-\frac{b^2}{32\hbar^2\omega^2}\left(\frac{eE_0}{\hbar\omega}\right)^4,
\end{equation}
\begin{equation}\label{ay}
\frac{A_y}{a}=1-\frac{a^2}{2\hbar^2\omega^2}\left(\frac{eE_0}{\hbar\omega}\right)^2+\frac{15b^2}{32\hbar^2\omega^2}\left(\frac{eE_0}{\hbar\omega}\right)^4,
\end{equation}
\begin{equation}\label{bx}
\frac{B_x}{b}=1-\frac{a^2}{2\hbar^2\omega^2}\left(\frac{eE_0}{\hbar\omega}\right)^2+\frac{b^2}{32\hbar^2\omega^2}\left(\frac{eE_0}{\hbar\omega}\right)^4,
\end{equation}
\begin{equation}\label{by}
\frac{B_y}{b}=1-\frac{4b\delta_0-3a^2}{2\hbar^2\omega^2}\left(\frac{eE_0}{\hbar\omega}\right)^2+\frac{33b^2}{32\hbar^2\omega^2}\left(\frac{eE_0}{\hbar\omega}\right)^4.
\end{equation}

\section{$T$ symmetry and circularly polarized light}\label{appendix:b}

It is clear that $h_\mathbf{-k}^\ast=h_\mathbf{k}^T$ in (\ref{inham}), and one can easily show that $H$ respects time-reversal symmetry. In fact, in this case $t\rightarrow-t$, $\mathbf{k}\rightarrow\mathbf{-k}$, and the transformed Hamiltonian
\begin{equation}\label{time}
\tilde{H}=\left[U_TH_\mathbf{-k}U^{-1}_T\right]^\ast, \quad U_T=\left(\begin{array}{cc}
0 & I \\
I & 0
\end{array}\right),
\end{equation}
where the asterisk stands for complex conjugation and $I$ is a $2\times2$ unity matrix. In the previous appendix we explicitly showed that the presence of off-resonant linearly polarized light results in the parameters of the Hamiltonian (\ref{inham}) being renormalized, such that $T$ symmetry is preserved for (\ref{effham}). The latter no longer holds if a system is driven by a circularly polarized laser, e.g.,
\begin{equation}
\mathbf{E}(t)=E_0\left[\mathbf{e}_y\cos(\omega t)-\mathbf{e}_x\sin(\omega t)\right],
\end{equation}
with amplitude $E_0$ and frequency $\omega$. Remarkably, doing the Peierls substitution selects only zeroth and first harmonics in the Hamiltonian (\ref{inham}),
\begin{equation}
H(t)=\sum\limits_{n=0,\pm1}\left(\begin{array}{cc}
h^{(n)}_\mathbf{k} & i\gamma\hat{\sigma}_x\delta_{n0} \\
-i\gamma\hat{\sigma}_x\delta_{n0} & \left[h^{(n)}_\mathbf{k}\right]^T
\end{array}\right)e^{in\omega t}=\sum\limits_{n=0,\pm1}H^{(n)}e^{in\omega t},
\end{equation}
with
\begin{equation}
h_\mathbf{k}^{(0)}=\left(\begin{array}{cc}
\delta_0-b\left(\frac{eE_0}{\hbar\omega}\right)^2-bk^2 & ia(k_x-ik_y) \\
-ia(k_x+ik_y) & -\delta_0+b\left(\frac{eE_0}{\hbar\omega}\right)^2+bk^2
\end{array}\right),
\end{equation}
and
\begin{equation}
h_\mathbf{k}^{(1)}=\frac{eE_0}{\hbar\omega}\left(\begin{array}{cc}
-b(k_x-ik_y) & 0 \\
-ia & b(k_x+ik_y)
\end{array}\right)=\left[h_\mathbf{k}^{(-1)}\right]^\dagger.
\end{equation}
In the framework of high-frequency expansion the effective Hamiltonian up to $1/\omega$ becomes
\begin{equation}
H_\mathrm{eff}=H^{(0)}+[H^{(1)},H^{(-1)}]/(\hbar\omega)+\ldots.
\end{equation}
One can easily verify that
\begin{equation}\label{lin}
[H^{(1)},H^{(-1)}]=\left(\begin{array}{cc}
[h_\mathbf{k}^{(1)},h_\mathbf{k}^{(1)\dagger}] & 0 \\
0 & -([h_\mathbf{k}^{(1)},h_\mathbf{k}^{(1)\dagger}])^T
\end{array}\right)
\end{equation}
and
\begin{equation}
[h_\mathbf{k}^{(1)},h_\mathbf{k}^{(1)\dagger}]=a\left(\frac{eE_0}{\hbar\omega}\right)^2\left(\begin{array}{cc}
-a & -2ib(k_x-ik_y) \\
2ib(k_x+ik_y) & a
\end{array}\right).
\end{equation}
In contrast to linearly polarized radiation that has been considered in this paper the appearance of the nonzero contribution due to (\ref{lin}) results in an effective Hamiltonian that explicitly breaks time-reversal symmetry (TRS) by violating (\ref{time}). In particular, for a three-dimensional topological insulator breaking TRS results in the gap opening in the spectrum of the surface states, thus destroying their topological protection \cite{chen}.

\section{Analytical results for edge states}\label{appendix:c}

We consider a half-infinite plane $x>0$ occupied by the topological material. Taking into account the fact that the edge state in the presence of electron-hole symmetry is a zero mode, we can write the following equation for the states with $k_y=0$
\begin{equation}\label{edge1}
\tilde{H}(-i\partial_x,0)\bm{\Psi}_\mathrm{edge}(x)=0.
\end{equation}
The solution to (\ref{edge1}) results in edge states, which are known to decay exponentially in the bulk, i.e., as $x\rightarrow\infty$.  These edge states are given by 
\begin{equation}\label{edge4}
\bm{\Psi}^{(1)}_\mathrm{edge}(x)=\frac{C}{\sqrt{L_y}}\left(e^{-\lambda_+x}-e^{-\lambda_-x}\right)
\left(\begin{array}{c}
1 \\ 1 \\ -1 \\ 1
\end{array}\right)
\end{equation}
and 
\begin{equation}\label{edge4a}
\bm{\Psi}^{(2)}_\mathrm{edge}(x)=\frac{C}{\sqrt{L_y}}\left(e^{-\lambda_+^\ast x}-e^{-\lambda_-^\ast x}\right)
\left(\begin{array}{c}
1 \\ 1 \\ 1 \\ -1
\end{array}\right),
\end{equation}
where we have defined normalization factor $C=\left[\frac{1}{2\mathrm{Re}\lambda_+}+\frac{1}{2\mathrm{Re}\lambda_-}-2\mathrm{Re}\left(\frac{1}{\lambda_++\lambda_-^\ast}\right)\right]^{-1/2}$ and
\begin{equation}\label{edge6}
\lambda_\pm=\frac{A_x\pm\sqrt{A_x^2-4B_x\left(\tilde{\delta}_0+i\tilde{\gamma}\right)}}{2B_x}.
\end{equation}
For experimentally relevant parameters of the quantum well two effective length scales can be defined, $\ell_1=-B_x/A_x$ and $\ell_2=-A_x/\tilde{\delta}_0$, as well as $k_0=\tilde{\gamma}/A_x$, so that $\lambda_1=-1/\ell_1$ and $\lambda_2=-1/\ell_2+ik_0$. If $k_y$ is nonzero yet small, the wave function of the edge states can be approximated by Eqs.~(\ref{edge4}) multiplied by complex phases $e^{ik_yy}$ corresponding to the propagation along the edges. The high-frequency Hamiltonian can be projected onto these states $\bm{\Psi}_\mathrm{edge}^{(1)}(x)e^{ik_yy}$ and $\bm{\Psi}_\mathrm{edge}^{(2)}(x)e^{ik_yy}$. Keeping only the linear terms in $k_y$, we derive an effective Hamiltonian of the edge states in the following form:
\begin{equation}\label{edge7}
H_\mathrm{edge}(k_y)=-\frac{A_x\tilde{\delta}_0k_y}{\tilde{\delta}^2+\tilde{\gamma}^2}\left(\begin{array}{cc}
0 & \tilde{\delta}_0+i\tilde{\gamma} \\
\tilde{\delta}_0-i\tilde{\gamma} & 0
\end{array}\right).
\end{equation}
This expression is valid in the regime when $\ell_1\ll\ell_2,1/k_0$. The Hamiltonian (\ref{edge7}) corresponds to massless Dirac particles with effective group velocity equal to
\begin{equation}\label{velocity}
v=-\frac{A_x\tilde{\delta}_0/\hbar}{\sqrt{\tilde{\delta}^2+\tilde{\gamma}^2}}.
\end{equation}
We note here that there exists the possibility to change the group velocity as a function of pump intensity, and it opens a unique way to manipulate the direction of the propagation of edge spin currents in the system optically, which could enable several new technologies.

\section{Details of numerical calculations}\label{appendix:d}

Equation~(2) in the main text corresponds to the bulk Hamiltonian dressed with an off-resonant linearly polarized electromagnetic radiation. To find the edge states for a finite strip in full agreement with \cite{shen}, the following steps must be performed:

\begin{enumerate}

\item Extend the calculation to the whole Brillouin zone according to $k_i \rightarrow \sin\left(k_i\Delta \right)/\Delta$, and $k_i^2 \rightarrow 2\left[1-\cos\left(k_i\Delta\right)\right]/\Delta^2$, where $\Delta$ is the lattice spacing and $i=x,~y$. % \\ \\

\item Perform the Fourier transformation along the finite axis (we call it the $v$-axis), so that $\sin\left(k_v a\right)\rightarrow i\hat{c}^\dag_{j+1}\hat{c}_j/2$ and $\cos\left(k_v a\right)\rightarrow \hat{c}^\dag_{j+1}\hat{c}_j/2$, where $k_v$ is \textbf{not} a good quantum number due to the finite size of the strip and $j$ runs over the lattice sites. For the terms in the Hamiltonian that are free from $k_v$, i.e., associated with the periodic directions, the substitution results in $g\left(k_u\right) \rightarrow g\left(k_u\right) \hat{c}^\dag_j \hat{c}_j$, where $g$ is an arbitrary function that has no terms related to the wave vector along the direction of the finite strip. % \\ \\

\item The tight-binding equation, at a particular lattice point, can be written as follows:
\begin{equation}
\left[\Gamma^\dag~\Phi_{j-1} + \mathcal{F}_\mathbf{k}\Phi_j+\Gamma~ \Phi_{j+1}\right]=E \Phi_j,
\end{equation}

where $\Phi$ is the full spinor. Electron and hole bands are designated with subscripts $e$ and $h$, respectively, while up and down arrows represent spin direction,
\begin{equation}
\Phi_j  = \left( \begin{array}{l}
{\phi _{ \uparrow ,e, j}}\\
{\phi _{ \uparrow ,h, j}}\\
{\phi _{ \downarrow ,e, j}}\\
{\phi _{ \downarrow ,h, j}}
\end{array} \right).
\end{equation}

\item Finally, terms from step 2 are equated accordingly to find the functions $\mathcal{F}_\mathbf{k}$, $\Gamma$, and $\Gamma^\dag$, comparing the terms with step 3. The Hamiltonian for a finite system with four points along the finite dimension reads
\begin{equation}
H_{16\times 16} = \left( {\begin{array}{*{20}{c}}
{\mathcal{F}_\mathbf{k}} & \Gamma & 0 & 0\\
{{\Gamma ^\dag }} & {\mathcal{F}_\mathbf{k}} & \Gamma &0\\
0 & {{\Gamma^\dag}} & {\mathcal{F}_\mathbf{k}} & \Gamma \\
0 & 0 & {{\Gamma ^\dag }} & {\mathcal{F}_\mathbf{k}}
\end{array}} \right),
\end{equation}
Note that the Dirichlet boundary condition has been employed in the above matrix.

\end{enumerate}
For all of our calculations, we have used $100$ lattice points along the finite axis of the strip, and convergence for the number of lattice points has been checked.

\section{Range of validity of high-frequency expansion}\label{appendix:e}

\begin{figure}[htpb]
\centering
\includegraphics[scale=0.27]{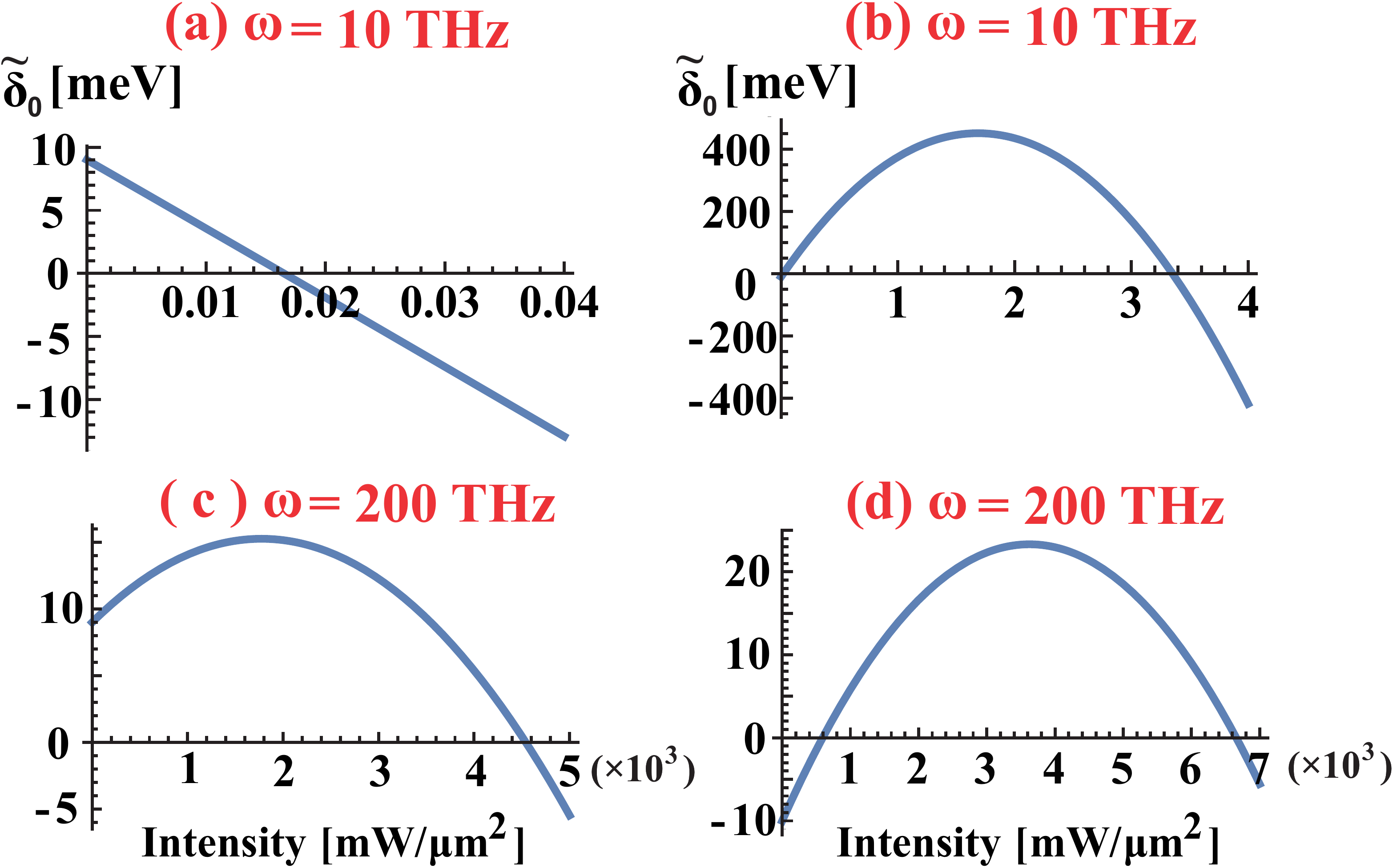}
\caption{Renormalization of the gap $\tilde{\delta}_0$ under an external pumping: starting from (a) the band insulator and (b) topological insulator at $\omega=10$ THz, as well as (c) the band insulator and (d) topological insulator at $\omega=200$ THz. Note the different energy scales for the four panels.}
\label{fig:fig}
\end{figure}

In this appendix we provide a physically motivated qualitative estimate on how large the frequency of a driving field should be to observe the predicted effects and present a rigorous mathematical formula. If we extend the original Hamiltonian (\ref{inham}), written in the vicinity of the $\Gamma$ point, to the whole Brillouin zone according to $k_i\rightarrow\sin(k_i\Delta)/\Delta$ and $k_i^2\rightarrow2\left[1-\cos(k_i\Delta)\right]/\Delta^2$, where $i=x,y$ and $\Delta$ stands for the lattice spacing, we can immediately see that the bandwidth of the system is determined by $\mathrm{max}\{a/\Delta,b/\Delta^2\}$. Putting $\Delta\sim 10\;\AA$, we find the bandwidth of the system to be of the order $0.1-1$ eV, which means that the frequency of an external driving $\omega\sim10^2-10^3$ THz, as pointed out in the main text. Meanwhile, our simulations reveal that the frequency $\omega$ may be lowered: in particular, we did not observe a qualitative discrepancy when raising the frequency up to 200 THz, whereas the approach in question becomes unreliable for lower frequencies, as shown in Fig.~\ref{fig:fig}, where two different frequencies (10 and 200 THz) of the dressing electromagnetic field are considered.

It is also instructive to derive a rigorous mathematical criterion to validate the use of high-frequency expansion. Consider the original Hamiltonian given by formula (\ref{inham}), and for simplicity neglect the contribution due to $\gamma$ as it has a rather small impact on the estimation of the valid frequency regime of the dressing electromagnetic field. Let us focus our attention on the block 
\begin{widetext}
\begin{equation}
G=\left(\begin{array}{cc}
A+B\cos(\omega t)+C\cos(2\omega t) & D+E\cos(\omega t) \\
D^\ast+E\cos(\omega t) & -A-B\cos(\omega t)-C\cos(2\omega t)
\end{array}\right).
\end{equation}
\end{widetext}
Assume $\Phi=(\phi_1,~\phi_2)$ is the solution to the dynamical problem $i\hbar\partial_t\Phi=G\Phi$, and with the help of the substitution
\begin{equation}
\Phi\rightarrow\Phi'=\exp\left\lbrace-i\sigma_z[2B\sin(\omega t)+C\sin(2\omega t)]/(2\hbar\omega)\right\rbrace\Phi,
\end{equation}
where $\sigma_z$ is the corresponding Pauli matrix, the matrix $G$ transforms according to
\begin{widetext}
\begin{equation}
G\rightarrow G'=\left(\begin{array}{cc}
A & f(t)e^{i[2b\sin(\omega t)+c\sin(2\omega t)]/(\hbar\omega)} \\
f^\ast(t) e^{-i[2b\sin(\omega t)+c\sin(2\omega t)]/(\hbar\omega)} & -A
\end{array}\right)
\end{equation},
\end{widetext}
where $f(t)=D+E\cos(\omega t)=\sum_mf_me^{im\omega t}$, provided that
\begin{equation}
f_m=\sum\limits_{n=-\infty}^\infty \left[DJ_{m-2n}\left(\frac{2b}{\hbar\omega}\right)+EJ_{m-2n+1}\left(\frac{2b}{\hbar\omega}\right)\right]J_n\left(\frac{c}{\hbar\omega}\right),
\end{equation}
where $J_n(z)$ is the $n$th-order Bessel function of the first kind. For a dynamical problem the solution can be tried as $\Phi'=e^{-i\varepsilon t/\hbar}\sum_n\Phi'_ne^{in\omega t}$, with the quasiparticle energy $\varepsilon$. Thus, the dynamical problem is equivalent to the stationary one in an extended Hilbert space,
\begin{subequations}
\begin{equation}
\varepsilon\Phi'_m=\sum_nG'_{mn}\Phi_n',
\end{equation}
\begin{equation}
G'_{mn}=\left(\begin{array}{cc}
\left(A+m\hbar\omega\right)\delta_{mn} & f_{m-n} \\
f_{n-m}^\ast & \left(-A+m\hbar\omega\right)\delta_{mn}
\end{array}\right).
\end{equation}
\end{subequations}
Interestingly, the projection to zero-photon subspace is equivalent to neglecting off-diagonal terms in $G'_{mn}$; the latter is achieved for any $m$ as long as

\begin{equation}
\left\vert\frac{f_m}{2A-m\hbar\omega}\right\vert\sim\mathrm{max}\{\vert D\vert,\vert E\vert\}\left\vert\frac{J_m\left(\frac{C+2B}{\hbar\omega}\right)}{2A-m\hbar\omega}\right\vert\ll 1.
\end{equation}
If one uses the parameters of the original Hamiltonian the latter can be rewritten as follows:
\begin{equation}
\frac{\mathrm{max}\{ak,a\lambda\}}{\left\vert 2\delta_0+b\lambda^2-2bk^2-m\hbar\omega\right\vert}\ll 1,
\end{equation}
where $\lambda=eE_0/(\hbar\omega)$ and we made allowance for the fact that $\vert J_m(z)\vert\leq 1$, thus providing the rigorous criterion for high-frequency expansion to be carried out.

\end{document}